\documentclass[usenatbib]{aa}  

\usepackage{natbib,twoopt}
\usepackage{graphicx}
\usepackage{xcolor}
\usepackage{txfonts}
\usepackage{academicons}
\usepackage{hyperref}
\usepackage{cleveref}
\usepackage{xcolor}
\usepackage[T1]{fontenc}

\usepackage[normalem]{ulem}

\newcommand{\hmpcinv}[0]{h{\rm Mpc}^{-1}}
\newcommand{\hinvmpc}[0]{h^{-1}{\rm Mpc}}
\newcommand{\nd}{n_{\rm d}}

\newcommand{\Pm}{P_{\rm m}}
\newcommand{\ellp}{\ell^{\prime}}
\newcommand{\kp}{k^{\prime}}
\newcommand{\Om}{\Omega_{\rm m}}
\newcommand{\Msun}{M_{\odot}}

\newcommand{\de}{\ensuremath{\mathrm{d}}}
\newcommand{\true}{^{\rm true}}
\newcommand{\ob}{^{\rm ob}}

\begin{document}

   \title{Power spectrum of the CODEX clusters}

   \subtitle{}

   \author{V. Lindholm\inst{1,2}\thanks{\email{valtteri.lindholm@helsinki.fi}}
          \and
          A. Finoguenov\inst{1}
          \and
          A. Balaguera-Antolínez\inst{3,4}
          \and
          T. Castro\inst{5,6,7}
          }

   \institute{Department of Physics, P.O. Box 64, 00014 University of Helsinki, Finland
        \and
        Helsinki Institute of Physics, Gustaf H{\"a}llstr{\"o}min katu 2, University of Helsinki, Helsinki, Finland
        \and
        Instituto de Astrof\'{\i}sica de Canarias, s/n, E-38205, La Laguna, Tenerife, Spain 
        \and
        Departamento de Astrof\'{\i}sica, Universidad de La Laguna, E-38206, La Laguna, Tenerife, Spain 
        \and
        INAF-Osservatorio Astronomico di Trieste, Via G. B. Tiepolo 11, 34143 Trieste, Italy 
        \and
        INFN, Sezione di Trieste, Via Valerio 2, 34127 Trieste TS, Italy
        \and
        IFPU, Institute for Fundamental Physics of the Universe, Via Beirut 2, 34151 Trieste, Italy }

   \date{Received December 11, 2024; accepted March 19, 2025}

  \abstract
   {}
   {We analyze the clustering of galaxy clusters in a large contiguous sample, the Constrain Dark Energy with X-ray (CODEX) sample. We construct a likelihood for cosmological parameters by comparing the measured clustering signal and a theoretical prediction, and use this to obtain parameter constraints.}
   {We measured the three multipole moments (monopole, quadrupole, and hexadecapole, $\ell = 0, 2, 4$) of the power spectrum of a subset of the CODEX clusters. To fully model cluster clustering, we also determined the expected clustering bias of the sample using estimates for the cluster masses and a mass-to-bias model calibrated using $N$-body simulations. We estimated the covariance matrix of the measured power spectrum multipoles using a set of simulated dark-matter halo catalogs. Combining all these ingredients, we performed a Markov chain Monte Carlo sampling of cosmological parameters $\Om$ and $\sigma_8$ to obtain their posterior.}
   {We found the CODEX clustering signal to be consistent with an earlier X-ray selected cluster sample, the REFLEX II sample. We also found that the measured power spectrum multipoles are compatible with the predicted, bias-scaled linear matter power spectrum when  the cosmological parameters determined by the Planck satellite are assumed. Furthermore, we found the marginalized parameter constraints of $\Om = 0.24_{-0.04}^{+0.06}$ and $\sigma_8 = 1.13_{-0.24}^{+0.43}$. The full 2D posterior is consistent, for example, with the Planck cosmology within the 68\% confidence region.}
   {}

   \keywords{astrophysics --
                cosmology --
                galaxy clusters
               }

   \maketitle

\section{Introduction}
    The clustering of galaxy clusters is sensitive to the geometry and the growth of structure in the Universe. Because of this, it has been an active field of cosmological research \citep[see e.g.,][]{1995lssu.conf..209B,borgani:1999, moscardini:2000, estrada:2009, Snchez2005CrosscorrelationsOX, 2011ARA&A..49..409A, marulli:2018, marulli:2021, moresco:2021, garrel:2022, lesci:2022, fumagalli24}. X-ray selected galaxy clusters, in particular, have been shown to possess great potential as a cosmological probe \citep[see e.g.,][]{1999Msngr..95...27G, 2003A&A...398..867S, 2005RvMA...18...76S,2009A&A...498..361P, Guzzo2009TheRG,2012MNRAS.422...44P,2014A&A...570A..31B,2014MNRAS.440.2077M,2017MNRAS.469.3738S,2019A&A...628A..43K,clerc:2023,2024A&A...689A.298G}. This paper is another contribution to this field of research.

    The dataset we analyze in this paper is the Constrain Dark Energy with X-ray (CODEX) galaxy cluster catalog \citep{finoguenov:2020}. This is a large catalog of X-ray sources identified as galaxy clusters by detecting an optical counterpart. The combined X-ray and optical observations allow for an accurate characterization of cluster properties, such as their masses. We give a description of the CODEX catalog and our sample selection in Section~\ref{sec:codex}.

    This paper aims to present the measurements of the power spectrum of the CODEX sample and infer cosmological information in the form of constraints on the cosmological parameters. Previous measurements \citep[][]{2001A&A...368...86S,2010MNRAS.401.2477H,2011MNRAS.413..386B} have characterized the cluster power spectrum as a scaled version of the underlying dark matter power spectrum, and show how the cluster clustering amplitude depends on observed properties such as the X-ray luminosity and richness, which paves the way toward a joint astrophysical and cosmological analysis \citep[][]{2011ARA&A..49..409A,2014MNRAS.441.3562E,2014A&A...563A.141B}.  In Section~\ref{sec:power_spectrum} we describe the respective methods to measure the cluster power spectrum and to model it. We present our measurements of the CODEX power spectrum and the results of our cosmological analysis in Section~\ref{sec:results}. Section~\ref{sec:conclusions} provides the conclusion of our analysis and results.

    This paper is a continuation of the work presented in \citet{lindholm:2021} (L21 hereafter), where the clustering of CODEX clusters was studied using the two-point correlation function. Theoretically, the two-point correlation function is simply the Fourier transform of the power spectrum, so the quantities are equivalent. In practice, however, the two quantities rely on completely different estimators and are thus complementary. In addition to measuring different clustering statistics, this work includes a few improvements to the L21 analysis. The most important ones are a more accurate covariance matrix (Section~\ref{sec:simulations}) and a proper handling of uncertainty in the cluster mass estimates (Section~\ref{sec:results}).

\section{The CODEX catalog}
    \label{sec:codex}
    The CODEX galaxy cluster survey is constructed by applying the red-sequence matched-filter probabilistic percolation (redMaPPer) algorithm \citep{rykoff:2014} to the Sloan Digital Sky Survey (SDSS, \citealp{blanton:2017}) photometry inside the $10\ 000$ square degree area of the Baryon Oscillation Spectroscopic Survey (BOSS, \citealp{dawson:2013}) footprint and identifying faint X-ray sources detected in the ROentgen SATellite (ROSAT) All-Sky Survey (RASS) \citep{tumper:1993, voges:1999}. A detailed description of the survey and the catalog is presented in \cite{finoguenov:2020}.

    In what follows, we refer to several probability distributions that are logarithmically normal. Thus, we define the following quantities: $r_c\equiv\ln (R_c/\rm{kpc})$ (core radius of the X-ray surface brightness), $l\equiv\ln (L_x/\rm{ergs}/s)$ (rest-frame X-ray luminosity in the 0.1--2.4 keV band), $\mu\equiv\ln (M_{200c}/\Msun)$ (total mass measured within the overdensity of 200 with respect to the critical density), $\lambda\equiv\ln (\mathrm{SDSS\; Richness})$ (defined at the optical peak, with a detailed description provided in \citealp{rykoff:2014}).
    
    To perform the clustering analysis, we selected the part of the CODEX catalog characterized by a low probability of chance cluster identification. Following previous CODEX studies, we applied the following cut, based on the measurement of SDSS richness,
    \begin{equation}
        \exp(\lambda) > 22(z/0.15)^{0.8}, 
    \end{equation}
    where $z$ is the redshift of the cluster. Further discussion of cleaning is presented in \citet{klein19}. We describe the effect of this cleaning as a $P^\mathrm{RASS}(I | \lambda, z)$ term in the modeling. As well as applying a redshift-dependent richness cut, we also excluded all clusters with a richness below 25.
     
    We applied the BOSS stellar mask to remove the areas in which stars affect optical cluster detection. We assume that the optical completeness of the CODEX catalog, above the applied richness cut, is constant across the BOSS area and model it using
    \begin{equation}
    \lambda_{50\%}(z) = \ln (17.2 + e^{\left( \frac{z}{0.32} \right)^{2}})
    .\end{equation}
    This is obtained using the tabulations of \citep{rykoff:2014}. We use an error function with the mean of $\lambda_{50\%}(z)$ and a $\sigma=0.2$, which reproduces the 75\% and 90\% quantiles of the distribution tabulated in \citep{rykoff:2014}. The probability of optical detection of a cluster in SDSS data is modeled as
    \begin{equation}
    P^\mathrm{SDSS}(I|\lambda,z)=1-0.5\mathrm{erfc}\left(\frac{\lambda-\lambda_{50\%}}{0.2\sqrt{2}}\right)
    \label{eq.psdss}
    .\end{equation}

    The RASS survey coverage is highly inhomogeneous, with the limiting flux varying by an order of magnitude. To properly account for the variations in the cluster distribution caused by this, we generate a random catalog, with a total number of objects six times that of the data catalog. We partition the survey area into 100 zones of equal sensitivity, $S$, with each consecutive zone having a 12\% difference in flux sensitivity. We denote the sky area of these zones as $\Delta\Omega_{S}$. The probability of cluster detection is computed as
    \begin{multline}
    \label{eqn:matrix}
    P(I|S,\mu,z,\nu) 
    = 
    \iiint \de l\true \de r_c \de \eta\ob P(I | \eta\ob , \beta(\mu), r_c )\\
    P(\eta\ob | \eta\true(l\true, S, z))  P(r_c, l\true | \mu, \nu, z),
    \end{multline}
    where $\eta$ denotes the X-ray count (superscript ``true'' stands for predicted, ``ob'' for detected), $\nu\equiv\frac{\lambda - \langle\lambda | \mu,z\rangle}{\sigma_{\lambda | \mu}}$, and all the probability distributions are described in detail in \cite{finoguenov:2020}. Equation~\ref{eqn:matrix} takes into account how the X-ray shapes ($r_c$, $\beta$) of the clusters and the RASS sensitivity affect the selection of the clusters, and it predicts changes in the distribution of X-ray shapes based on the measured covariance of the properties of the cluster \citep{cavaliere:1976, mulroy:2019, farahi:2019, kaefer:2019}. This approach allows us to account for the anticorrelation between X-ray luminosity scatter and optical richness, as well as the anticorrelation between galaxy cluster core radii and their luminosity scatter. Finally, we estimate the expected number of clusters in each redshift bin as
    \begin{equation}
    \label{eqn:NC}
    \langle N(\Delta z) \rangle = \Delta\Omega_{S} \int_{\Delta z} \de z  \frac{\de V}{\de z \de \Omega}  (z) 
    \iint \de \mu \de \lambda \frac{\de n(\mu,\lambda, S, z)}{\de \mu \de \lambda \de V}
    ,\end{equation} 
    where
    \begin{multline}
    \label{eqn:hmf}
    \frac{\de n(\mu,\lambda, S, z)}{\de \mu \de \lambda \de V}
    = P^\mathrm{RASS}(I | \lambda, z)P^\mathrm{SDSS}(I | \lambda, z) \\
    P(I|S,\mu,z,\nu) P(\lambda| \nu, \mu ) \frac{\de n(\mu, z) }{\de V \de\mu} \,
    .\end{multline}
    
\begin{figure}     
        \centering
        \includegraphics[width=\hsize]{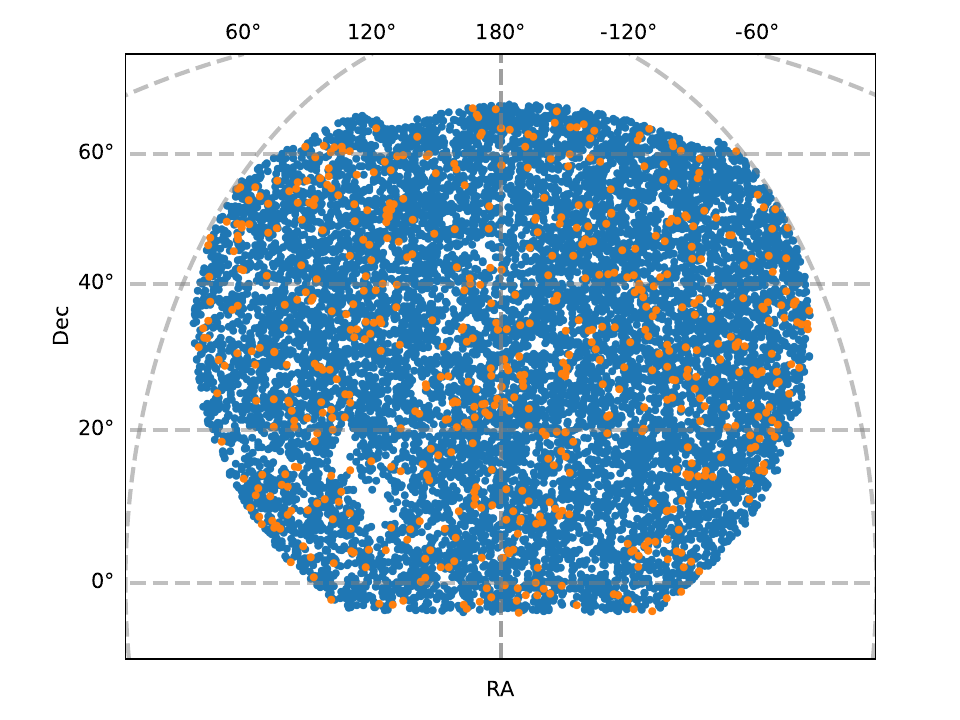}
        \caption{Sky footprint of the subset of the CODEX catalog used for computing the power spectrum. The plot is a zoomed-in view of a Mollweide projection. The orange points show the clusters and the blue points show the random points.}
        \label{fig:footprint_codex}
    \end{figure}

\begin{figure}
        \centering
        \includegraphics[width=\hsize]{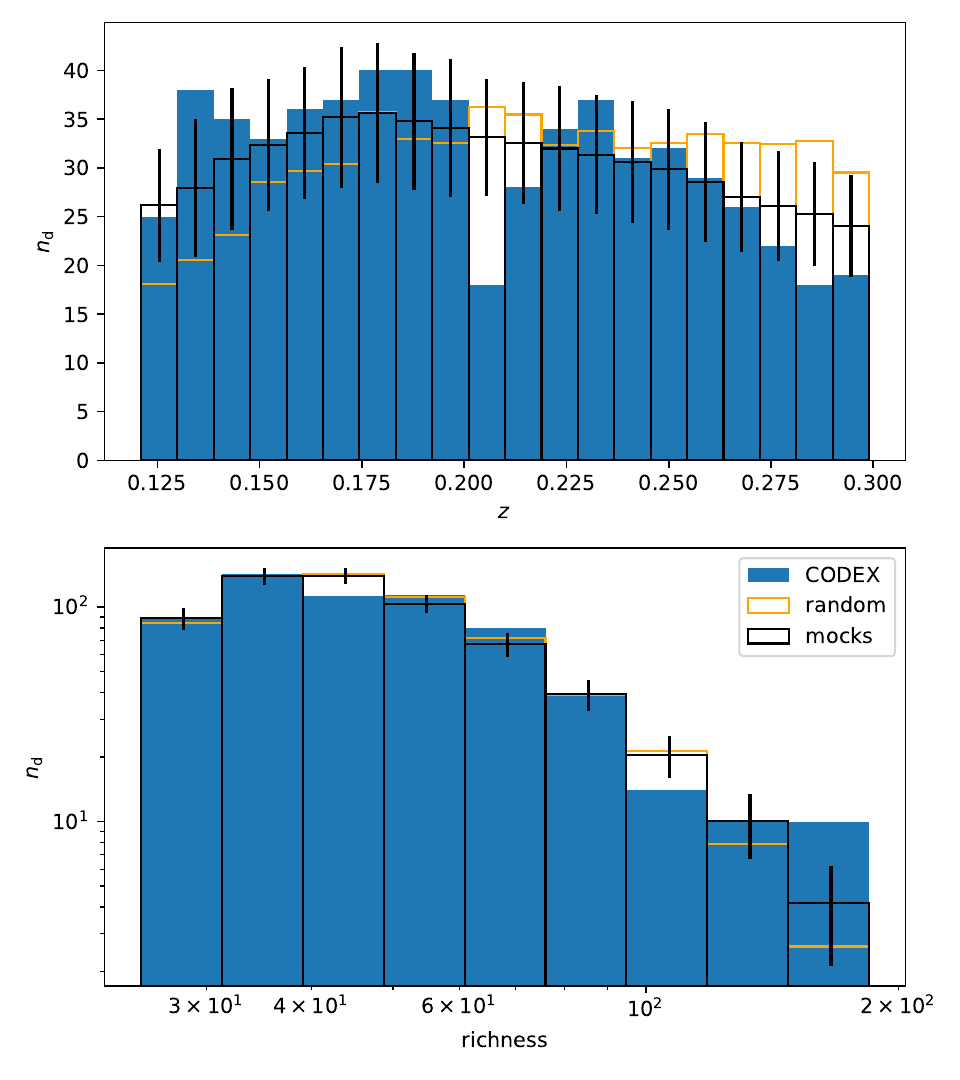}
        \caption{One-point statistics of the random catalog and the mock catalogs compared to the corresponding CODEX quantities. The blue bars show the CODEX sample and the empty orange bars show the random sample. The bars that correspond to the random sample have been normalized with the ratio of objects in the CODEX and random samples. The empty black bars show the mean values over all the mock catalogs and the error bars show the standard deviation. Top panel: Redshift. Bottom panel: Richness.}
        \label{fig:1p_stats}
   \end{figure}

    The sample selection in terms of richness and the construction of the random catalog described above is the same as in L21. However, our final sample differs from L21 in two ways. First, we selected a subset of the full CODEX footprint to have the sky area fully covered by simulated catalogs. The CODEX catalog consists of two disjoint patches, one of which has slightly larger angular coverage than the mock catalogs available to us (see Section~\ref{sec:simulations}). We thus selected a subset of the larger patch to be able to cover the whole footprint with one set of mock catalogs. Second, we also used a slightly narrower redshift range, namely $z=0.12-0.30$ (as opposed to $z=0.1-0.5$), to be compatible with the simulations. The final cluster sample we used for the power spectrum estimation contains 615 objects and the corresponding random sample is 11160 objects. The sky footprint of each sample can be seen in Fig.~\ref{fig:footprint_codex} and the corresponding redshift and richness distributions in Fig.~\ref{fig:1p_stats} (along with those of the simulated mock catalogs). We also show the standard deviation of the mock catalogs as error bars. Clearly, the differences in the distributions are within the estimated variance, so the three types of catalogs (CODEX, random, and mocks) are compatible with being drawn from the same ensemble.

\section{Cluster power spectrum: measurements and modeling}
    \label{sec:power_spectrum}
\subsection{Measurements}
    We performed the measurements of a three-dimensional power spectrum in redshift space, $P_{\ell}(k)$, using the estimator of \citet[][]{2015MNRAS.453L..11B}, which is a generalization of the celebrated Feldman-Kaiser-Peacock (FKP) estimator \citep[][]{feldman:1994} applied to samples covering a wide area, in which the distant observer approximation is no longer valid. The minimum variance weights provided by the FKP estimator $w_{FKP}(\mathbf{r})=(1+P_{0}\bar{n}(\mathbf{r}))^{-1}$ were implemented using a power spectrum amplitude $P_0 = 1.5 \times 10^5(\hmpcinv)^3$ and the mean cluster density $\bar{n}(\mathbf{r})$ was evaluated at each cluster position. The normalization and (Poisson) shot-noise contribution were estimated from the set of random tracers \citep[see e.g.,][for a detailed implementation of the estimator]{2011MNRAS.413..386B,beutler:2014}. The transformation from redshifts to comoving distances was performed under the assumption of a $\Lambda$CDM cosmological model with the parameters from \citet{planck:2018} (Planck 2018 cosmology hereafter). We estimated the CODEX power spectrum multipoles in 35 linearly spaced $k$ bins from $0.1 \hmpcinv$ to $0.35 \hmpcinv$. We used the \verb|nbodykit|\footnote{\url{https://github.com/bccp/nbodykit}} Python library \citep{hand:2018} to obtain estimates of the monopole ($\ell=0$), quadrupole ($\ell=2$), and hexadecapole($\ell=4$). The box used to perform the Fourier transform had sides with a comoving length of $(1133 \hinvmpc, 1151 \hinvmpc, 651 \hinvmpc)$ and the mesh resolution was 128 cells per side.

\subsection{Modeling}
    All the cosmology-dependent quantities mentioned in the following section were computed using the \verb|COLOSSUS|\footnote{\url{https://bitbucket.org/bdiemer/colossus/src/master/}} Python library \citep{diemer:2018}.
    
    We model the cluster power spectrum based on a linear matter power spectrum, which was computed using the transfer function of \citet{eisenstein:1998} (EH hereafter). Even though we tested more accurate models computed from Boltzmann solvers \citep[see e.g.][]{2011arXiv1104.2932L}, the differences are small compared to the statistical errors of the power spectrum. Therefore, we adopted the EH parameterization, which helps to increase the speed of the likelihood calculations (Section~\ref{sec:results}). We model the redshift-space cluster power spectrum using the Kaiser approximation \citep[][]{kaiser:1987}, which, for the three explored multipoles, is written as
    \begin{align}
            P_0(k) &= \left(1 + \frac{2}{3}\beta + \frac{1}{5}\beta^2\right)b^2\Pm(k),\label{eq:p0} \\
            P_2(k) &= \left(\frac{4}{3}\beta + \frac{4}{7}\beta^2 \right)b^2\Pm(k), \label{eq:p2} \\
            P_4(k) &= \frac{8}{35}\beta^2b^2\Pm(k). \label{eq:p4}
    \end{align}
    Here $\beta \equiv f/b$, where $f \equiv -{\rm d}\ln{D(z)}/{\rm d}\ln{(1+z)}$ is the growth rate, $b$ is the large-scale (or effective) galaxy cluster bias, and $\Pm(k)$ is the isotropic matter power spectrum.

     To obtain an estimate of the large-scale galaxy cluster bias, we follow the procedure of L21. First, we estimate the cluster masses using their observed richness and the scaling relation calibrated in \citet{kiiveri:2021} (K21 hereafter). The relation is a power law, parametrized as
     \begin{equation}
        \label{eq:scaling}
         \ln{\lambda_i} = \alpha\ln{(M_i/M_{\rm Piv})} + \beta.
     \end{equation}
     Here $\lambda_i$ is the richness and $M_i$ the mass of the $i$th cluster, $M_{\rm Piv}=10^{14.81}\Msun$ is a pivot mass, and $\alpha$ and $\beta$ are the model parameters to be calibrated. The particular calibration we used employs parameter priors from the South Pole Telescope Polarimeter (SPTpol) Extended Cluster Survey \citep{bleem:2020}. Equation~\ref{eq:scaling} can be inverted to estimate the mass of a cluster with an observed richness of $\lambda_i$. With these mass estimates, we compute a large-scale bias using the model calibrated in \cite{comparat:2017}\footnote{Note that this assigns the average bias to each halo, which is already averaged from the full sample, \citep[see e.g.,][]{2023arXiv231112991B}} An extensive comparison of various scaling relations and bias models in the context of the CODEX clusters can be found in L21. Finally, we compute a weighted mean over all the clusters in our sample:
     \begin{equation}
          \label{eq:mean_bias}
         \overline{b} = \frac{1}{\nd} \sum_{i=1}^{\nd} b(M_i, z_i)g(z_i).
     \end{equation}
    In this expression, $\nd$ is the number of clusters and $b(M_i, z_i)$ is the effective bias that corresponds to a cluster of mass $M_i$ at redshift $z_i$; $g(z_i) \equiv D(z_i)/D(0)$, where $D(z)$ is the growth function at redshift $z$. The growth is included to scale all the biases with respect to the matter power spectrum at $z = 0$. We use the bias factor of Eq.~(\ref{eq:mean_bias}) to describe the cluster power spectrum in Eqs.~(\ref{eq:p0})-(\ref{eq:p4}).

    The parameters that define the richness-mass scaling relation have a certain uncertainty, and there is also an intrinsic scatter of true cluster masses around the value predicted by the scaling relation. We take both of these effects into account as additional uncertainty when we fit the cosmological parameters. Details can be found in Section~\ref{sec:results}.

\subsection{Window function}
    \label{subsec:window}
    The measurements of the power spectrum using the aforementioned estimators are the response of the convolution between the survey window function (the Fourier transform of the selection function) and the underlying (theoretical) cluster power spectrum \citep{feldman:1994}. In a likelihood analysis, instead of deconvolving the measurements to obtain the underlying power spectrum (and retrieve cosmological information therefrom), it is simpler and numerically more stable to convolve a theoretical model $P^{\rm th}_{\ellp}(k)$  with the window function. Such a convolution can be transformed into a matrix multiplication of the form
    \begin{equation}
        P_{\ell}(k_i) = \sum_{\ellp=0,2,4}\sum_{j} W_{\ell\ellp}(k_i, \kp_j) P^{\rm th}_{\ellp}(\kp_j),
    \end{equation}
    where the mixing matrix $W_{\ell\ellp}(k_i, \kp_j)$ is computed using the approach of \citet[][]{beutler:2014}. We used 200 linearly spaced $\kp$ bins in the range $0.1\hmpcinv-0.35\hmpcinv$.
    
\begin{figure}
        \centering
        \includegraphics[width=\hsize]{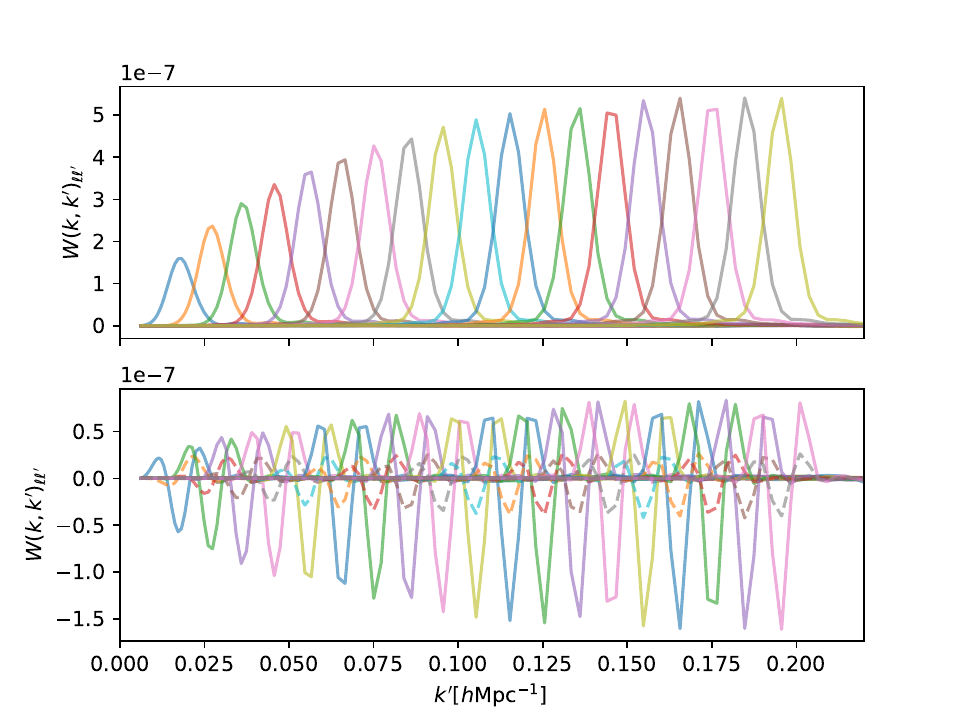}
        \caption{Elements of the matrix window that contribute to the power spectrum quadrupole. Top panel: The $\ell=2, \ell^{\prime}=2$ block. Bottom panel: The $\ell=2, \ellp=0$ block in solid lines and the $\ell=2, \ellp=4$ block in dashed lines. Each peak corresponds to a single wave number, $k_i$, at which the power spectrum multipoles are estimated.}
        \label{fig:matrix_window}
   \end{figure}

    As an example, we show three blocks of the estimated window function in Fig.~\ref{fig:matrix_window}. The first one is the quadrupole-quadrupole block $(\ell, \ellp) = (2, 2)$, the second one the quadrupole-monopole block $(\ell, \ellp) = (2, 0)$, and the third one the quadrupole-hexadecapole block $(\ell, \ellp) = (2, 4)$. The amplitude of the $(\ell, \ellp) = (2, 0)$ block is similar to the $(\ell, \ellp) = (2, 2)$ block, which highlights the anisotropy induced by the survey selection. The full window matrix is shown in the bottom panel of Fig.~\ref{fig:cov_codex}.

\subsection{Mock catalogs}
    \label{sec:simulations}
    To generate mock catalogs suitable for estimating the covariance matrix of the cluster power spectrum, we used a set of dark matter halo catalogs produced with the PINpointing Orbit Crossing Collapsed HIerarchical Objects (PINOCCHIO\footnote{\url{https://github.com/pigimonaco/Pinocchio}}) algorithm \citep{monaco:2002a, munari:2017}. These simulations correspond to the flat $\Lambda$CDM model, and the values for the cosmological parameters used in the simulation are listed in Table~\ref{tab:cosmo_pinocchio}. There are no massive neutrinos in the simulation.
    \begin{table}
    \caption{Values for the cosmological parameters used in the PINOCCHIO simulations.}
    \label{tab:cosmo_pinocchio}
    \centering
    \begin{tabular}{c c c c c c}
        \hline\hline
        $\Om$ & $\Omega_{\Lambda}$ & $\Omega_b$ & $h$ & $n_s$ & $\sigma_8$\\
        \hline
        0.3089 & 0.6911 & 0.0486 & 0.6774 & 0.9667 & 0.81 \\
        \hline
    \end{tabular}
  \end{table}
  
    We started with a set of 500 light cones that contain $\sim 6 \times 10^7$ dark matter halos. They cover a spherical cap of a radius of $45\deg$ and a redshift range of $z=0.0 - 1.0$. They are based on $1h^{-1}{\rm Gpc}$ simulation boxes with a $2048^3$ grid. The light cones are constructed with a redshift-dependent lower limit for halo mass. The limit has a maximum of $10^{13}h^{-1}\Msun$, which is well below the smallest cluster masses in our CODEX sample. We applied the CODEX selection function to the mock catalogs and assigned a cluster richness to each halo based on their masses. Finally, we applied the same redshift and richness cuts that define the CODEX sample to the resulting mock cluster catalog. Figure~\ref{fig:1p_stats} shows the redshift and richness distributions of the final mock samples, along with the CODEX sample. These demonstrate the compatibility between the mock catalogs and the CODEX catalog.

\begin{figure}
        \centering
        \includegraphics[width=\hsize]{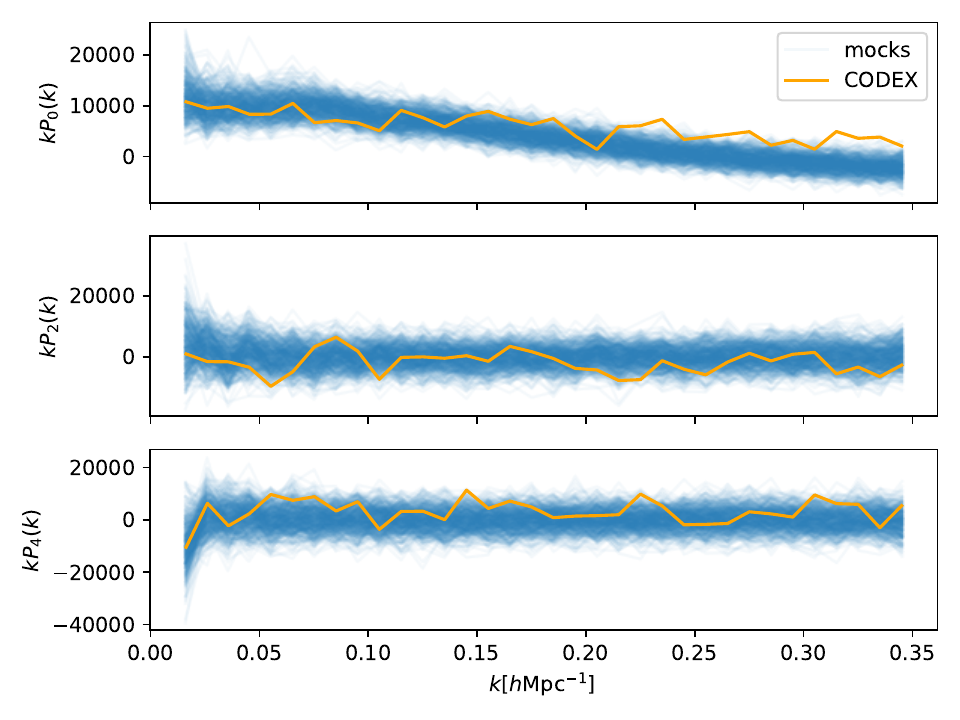}
        \caption{Power spectrum multipoles used to compute the covariance matrix. Top panel: Monopole. Middle panel: Quadrupole. Bottom panel: Hexadecapole. Each panel shows all the 500 simulated spectra used to compute the covariance in transparent blue. The orange line is the corresponding CODEX spectrum.}
        \label{fig:pk_set}
   \end{figure}

    In Fig.~\ref{fig:pk_set} we show all the simulated spectra used for computing the covariance matrix, along with the CODEX spectrum. The figure shows that the CODEX spectrum is compatible with the simulated ensemble up to wave numbers of $k \sim 0.2 \hmpcinv$ but has a larger amplitude for higher $k$. As shown by \citet{munari:2017} (Fig.~9, for example), the ability of the PINOCCHIO code to reproduce power spectra of $N$-body simulations starts to deteriorate at around these scales. Hence, we limited ourselves to scales $k < 0.2 \hmpcinv$ in our cosmological parameter estimation. This is because modeling the effects of non-linear growth of the structure that is not fully captured by the PINOCCHIO mock catalogs can lead to biased cosmological constraints \citep{munari17,fumagalli24} and is outside the scope of this work.

\begin{figure}
        \centering
        \includegraphics[width=\hsize]{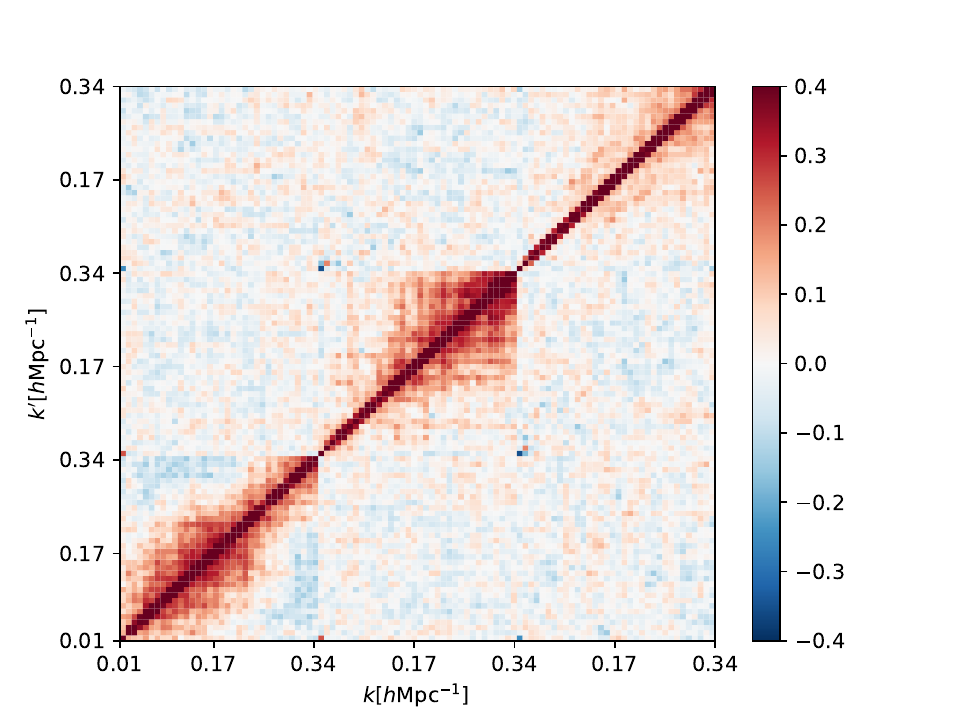}
        \includegraphics[width=\hsize]{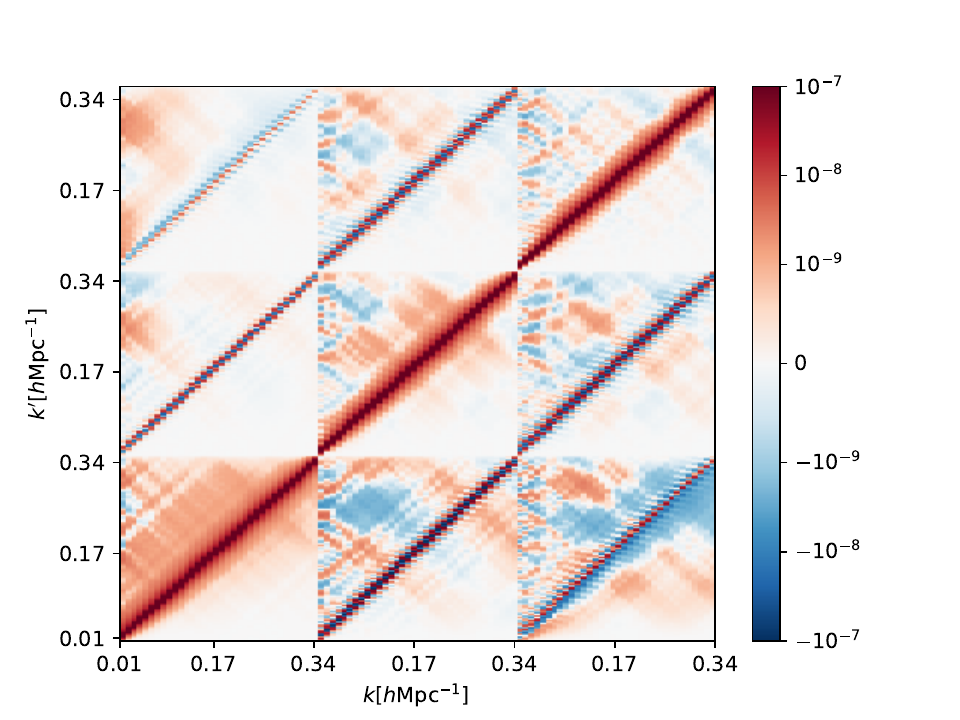}
        \caption{Top panel: Covariance matrix of the CODEX power spectrum multipoles, as computed from the PINOCCHIO mocks. The elements are normalized with the diagonal. Bottom panel: Full matrix window. Both matrices have the same structure. Starting from the bottom left corner, the diagonal blocks correspond to $\ell = 0, 2, 4$ auto-correlation. The off-diagonal blocks show the corresponding cross-correlations.}
        \label{fig:cov_codex}
   \end{figure}

    We compute the covariance matrix as the sample covariance
    \begin{equation}
        \label{eq:cov}
        C_{\ell \ellp}(k_i, k_j) = \frac{1}{n_{\rm m} - 1}\sum_{\alpha=1}^{n_{\rm m}} \left[ P^{\alpha}_{\ell}(k_i) - \overline{P}_{\ell}(k_i) \right]\left[P^{\alpha}_{\ellp}(k_j) - \overline{P}_{\ellp}(k_j) \right],
    \end{equation}
    where $P^{\alpha}_{\ell}(k_i)$ are the power spectrum multipoles computed from each mock, $\overline{P}_{\ell}(k_i)$ is their mean, and $n_{\rm m}$ is the number of mock catalogs. Figure~\ref{fig:cov_codex} shows the elements of the covariance matrix $C_{ij}$ normalized by the corresponding diagonal elements:
    \begin{equation}
        \rho_{ij} \equiv \frac{C_{ij}}{\sqrt{C_{ii}C_{jj}}},
    \end{equation}
    where indices $i,j$ are also taken to include the multipole numbers $\ell = 0, 2, 4$

\section{Results}
    \label{sec:results}
\begin{figure}
        \centering
        \includegraphics[width=\hsize]{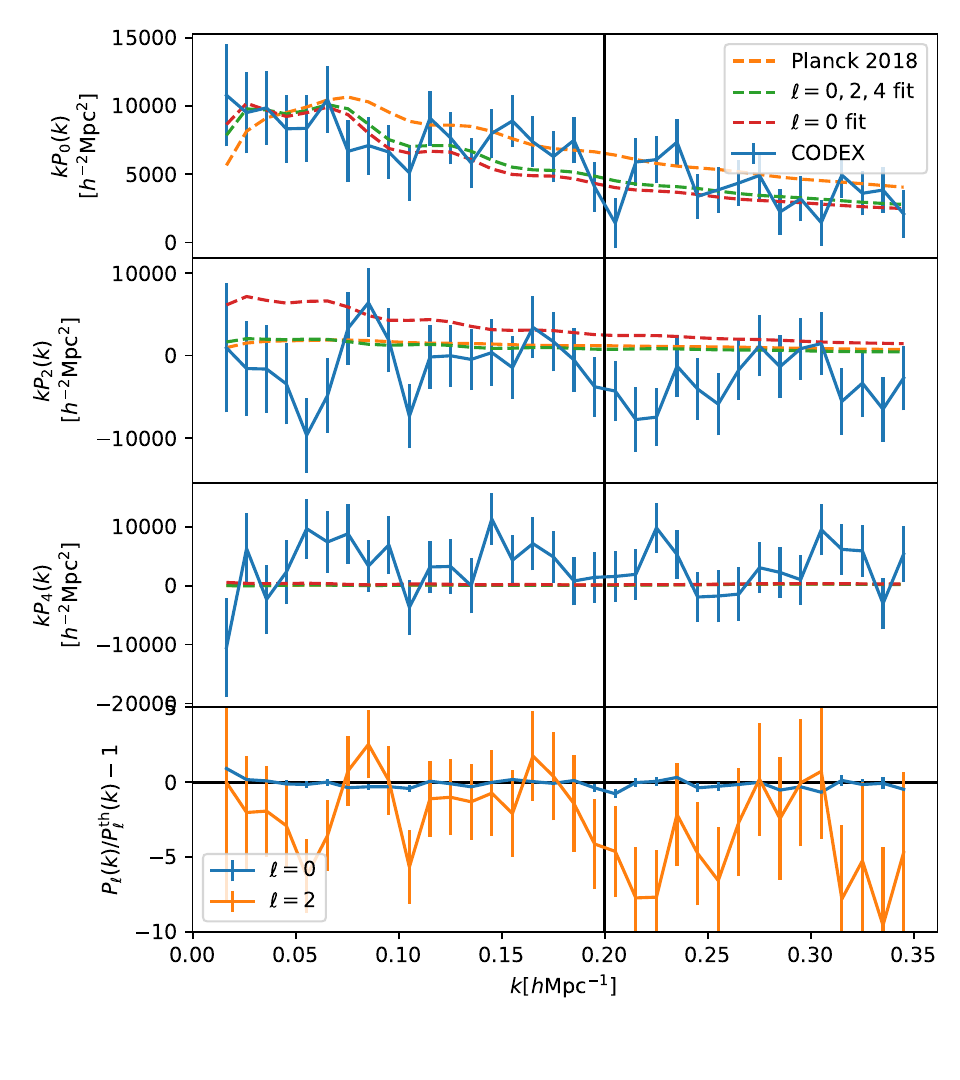}
        \caption{Multipoles of the CODEX cluster's power spectrum. The solid lines with error bars show the measurements, and the dashed lines show the corresponding theoretical predictions. The orange lines show the theoretical prediction using the Planck 2018 cosmology, the green lines the prediction using the best-fit cosmology obtained using all the multipoles, and the red lines the prediction using the best-fit cosmology obtained using the monopole alone. Top panel: Monopole ($\ell=0$). Second panel: Quadrupole ($\ell=2$) Third panel: Hexadecapole ($\ell=4$). Bottom panel: Relative difference of the measured and predicted spectra (Planck 2018 cosmology) for monopole and quadrupole. The predicted hexadecapole is close to zero, which makes the relative difference extremely large. The vertical black line shows up to which wavenumbers we include the measurements in our cosmological analysis.}
        \label{fig:pk}
    \end{figure}

\begin{figure}
        \centering
        \includegraphics[width=\hsize]{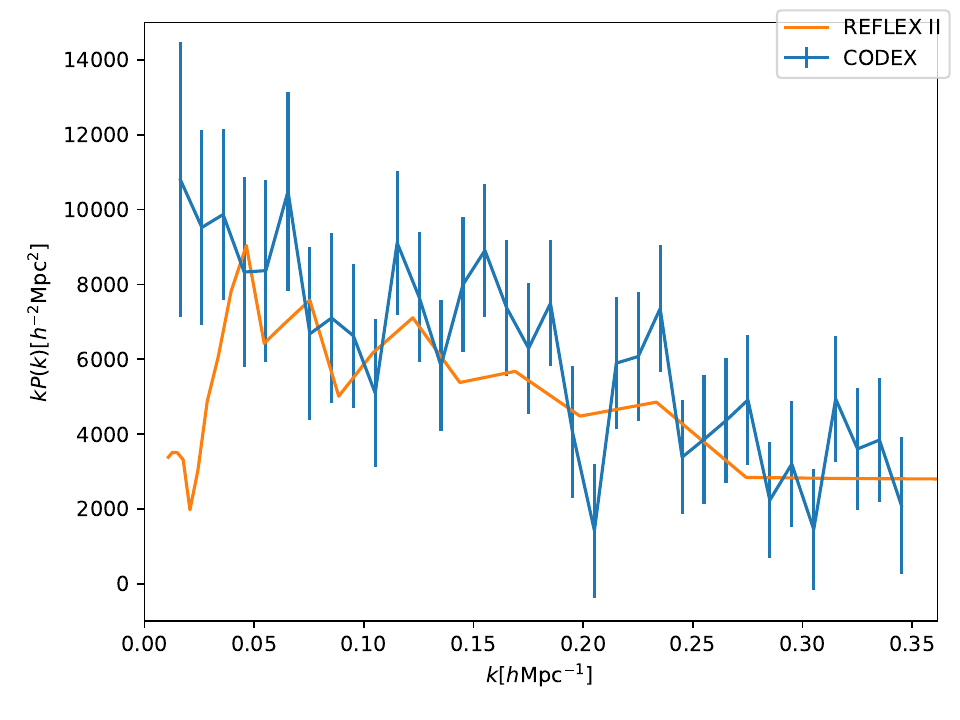}
        \caption{Comparison of power spectra from CODEX and REFLEX II samples. The blue line with error bars is the CODEX power spectrum monopole, and the orange line is the REFLEX II isotropic power spectrum.}
        \label{fig:pk_reflex}
    \end{figure}

    Figure~\ref{fig:pk} shows the power spectrum multipoles measured from the CODEX sample, along with the theoretical predictions from Eqs.~\ref{eq:p0}, \ref{eq:p2} and \ref{eq:p4}. We show predictions for the Planck 2018 cosmology, the  best-fit cosmology obtained by fitting all the CODEX power spectrum multipoles, and the best-fit cosmology obtained by fitting the monopole alone to highlight the importance of the higher multipoles in our cosmological analysis (see the following paragraphs for more details). In Fig.~\ref{fig:pk_reflex} we show a comparison of the power spectrum monopole measured from our CODEX sample (same as the solid blue line in the top panel of Fig.~\ref{fig:pk}) and the isotropic power spectrum measured from the REFLEX II sample \citep{2011MNRAS.413..386B}. In both figures, the error bars are computed as the square root of the diagonal elements of the covariance matrix of Eq.~\ref{eq:cov}. Our measurement of the CODEX power spectrum is in agreement with the Planck 2018 cosmology prediction within the error bars. A qualitative agreement with results from the REFLEX II samples is also seen, with differences likely being due to the redshift range and cluster mass cut used in each sample.
    
    The power spectrum measurements described above can be used to obtain constraints on cosmological parameters. To do this, we ran a Markov chain Monte Carlo (MCMC) sampling of the present-day matter density parameter $\Om$ and the power spectrum amplitude $\sigma_8$ to obtain their posterior. The other $\Lambda$CDM parameters were fixed to the Planck 2018 values. We expect the main uncertainty in our modeling to be associated with the cluster mass estimates. To fully account for this, we also sampled the two richness-to-mass scaling relation parameters $\alpha$ and $e^{\beta}$. We used Gaussian approximations to the posteriors presented in K21 as priors for these parameters. To include the scatter in the richness-to-mass conversion, we generated a Gaussian noise vector with $\sigma = 0.28$ (corresponding to the scatter determined in K21) and added this to the logarithmic mass of each cluster at each likelihood evaluation. The sampling was implemented using the \verb|emcee|\footnote{\url{https://github.com/dfm/emcee}} Python library \citep{foreman-mackey:2013}.
    
    We combine all the power spectrum multipoles into a single data vector, denoted with $\Vec{\hat{P}}$. The same goes for the covariance matrix, denoted with $\Vec{C}$. These are taken to be independent of the cosmology, even though estimating $\Vec{\hat{P}}$ and $\Vec{C}$ requires assuming a cosmology, to transform angles and redshifts into distances. Instead, we include the geometric effects of changing the cosmology in our modeling of the power spectrum multipoles following \citet{hector:2020}. The method consists of two steps:
    \begin{enumerate}
        \item Transforming the wave vector magnitude and the angle with the line of sight direction according to changes in the Hubble distance $D_{\rm H}(z) \equiv c/H(z)$ and the angular diameter distance due to differences in cosmology.
        \item Renormalizing all the multipoles with a single factor to account for the isotropic volume rescaling caused by the differences in cosmology.
    \end{enumerate}
    In our case, the reference cosmology is the one used to compute the CODEX power spectra and the covariance matrix (Planck 2018) and the comparison cosmology is the one used at each likelihood evaluation. Both transformations 1 \& 2 are redshift-dependent. Here we used the mean redshift of our sample, $z=0.20$. We assume the likelihood to be Gaussian in $\Vec{\hat{P}}$:
    \begin{equation}
        \ln{\mathcal{L(\Vec{\Theta})}} = -\frac{1}{2}\left[ \Vec{\hat{P}} -\Vec{P}(\Vec{\Theta})\right]^{\rm T} \Vec{C}^{-1} \left[ \Vec{\hat{P}} -\Vec{P}(\Vec{\Theta}) \right] + {\rm a\ constant}.
    \end{equation}
    Here vector $\Vec{\Theta}$ denotes the sampled parameters and $\Vec{P}(\Vec{\Theta})$ are the predicted power spectrum multipoles. The vector $\Vec{P}(\Vec{\Theta})$ depends on $\Om$ and $\sigma_8$ through both the linear matter power spectrum prediction and the mean bias of Eq~\ref{eq:mean_bias}. The dependence on the scaling parameters $\alpha$ and $e^{\beta}$ comes from the masses entering the computation of the mean bias. 
    
    We used the estimated power spectrum multipoles over scales of $k < 0.2 \hmpcinv$. Figure~\ref{fig:pk} suggests that the Planck 2018 cosmology model and the measurements agree up to the largest $k=0.35\hmpcinv$. We notice that, since our mock measurements exhibit a weaker monopole for $k \gtrsim 0.2 \hmpcinv$, the covariance matrix estimate might not be reliable on these scales. However, we did run a cosmological parameter fit including all scales up to $k=0.35\hmpcinv$ to assess their effect on the obtained parameter constraints. This is shown in Fig.~\ref{fig:mcmc_kmax} and the discussion at the end of this section.
    
\begin{figure}
        \centering
        \includegraphics[width=\hsize]{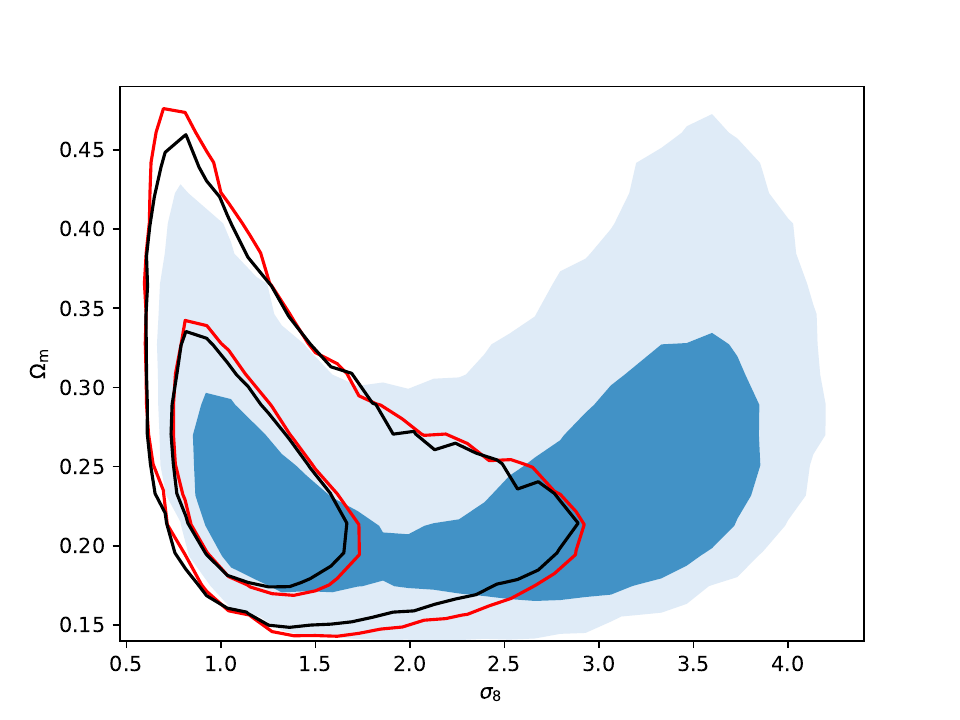}
        \caption{Posterior distribution for the two cosmological parameters $\Om$ and $\sigma_8$ in the case where varying sets of power spectrum multipoles are included in the sampling. The filled blue contours correspond to $\ell = 0$ only, the red contours correspond to $\ell = 0, 2$, and the black contours to $\ell = 0, 2, 4$. The contours correspond to the 68\% and 95\% confidence regions.}
        \label{fig:mcmc_multipoles}
    \end{figure}

\begin{figure}
        \centering
        \includegraphics[width=\hsize]{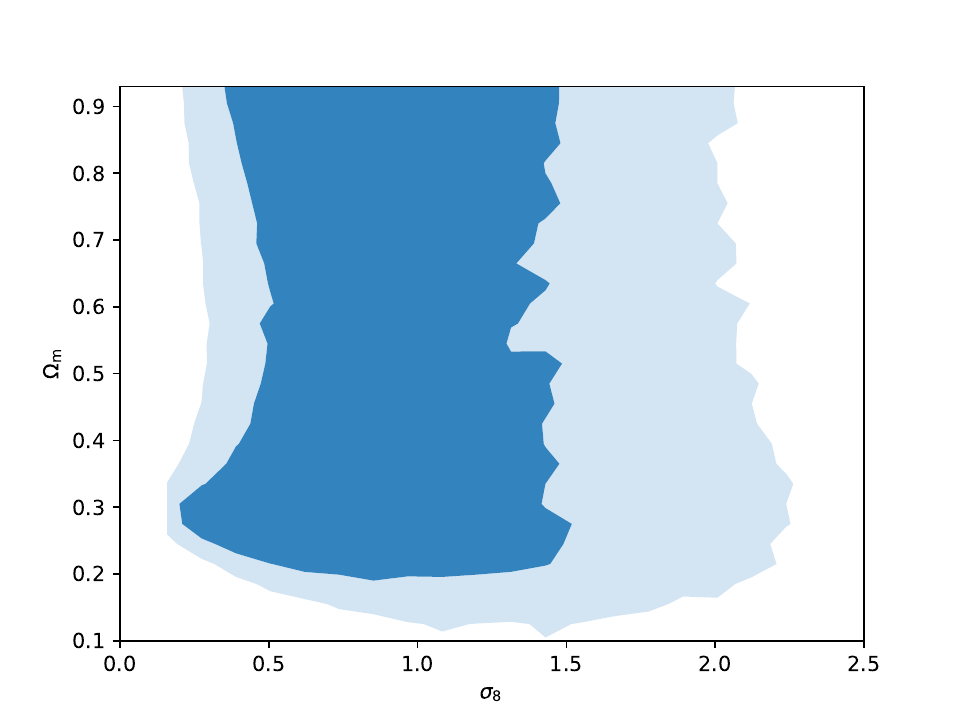}
        \caption{Posterior distribution for the two cosmological parameters $\Om$ and $\sigma_8$ in the case where only the power spectrum quadrupole is included in the sampling. The contours correspond to the 68\% and 95\% confidence regions.}
        \label{fig:mcmc_ell2}
    \end{figure}

    In Fig.\ref{fig:mcmc_multipoles} we show the 2D posterior distributions for $\Om$ and $\sigma_8$ in the case where $\ell = 0$, $\ell = 0,2$, and $\ell = 0,2,4$ are included in the data vector. For the monopole alone, we found the distribution to be bimodal, with one peak at around canonical values of $\sigma_8 \sim 1$ and another one with significantly larger values of $\sigma_8 \gtrsim 3$. Adding the quadrupole removes the latter peak due to the large amplitude of the model $P_2(k)$ in this region (see Fig.~\ref{fig:pk}), which is inconsistent with the measurement. This effect is also shown in Fig.~\ref{fig:mcmc_ell2}, in which we show the posterior distribution for $\Om$ and $\sigma_8$ when we only use the quadrupole. In this case, the peak at $\sigma_8 \gtrsim 3$ is excluded with high significance. Including the hexadecapole has a significantly smaller effect, but still decreases the area of the $68\%$ and $95\%$ confidence regions by $13\%$ and $6\%$, respectively.
    
\begin{figure}
        \centering
        \includegraphics[width=\hsize]{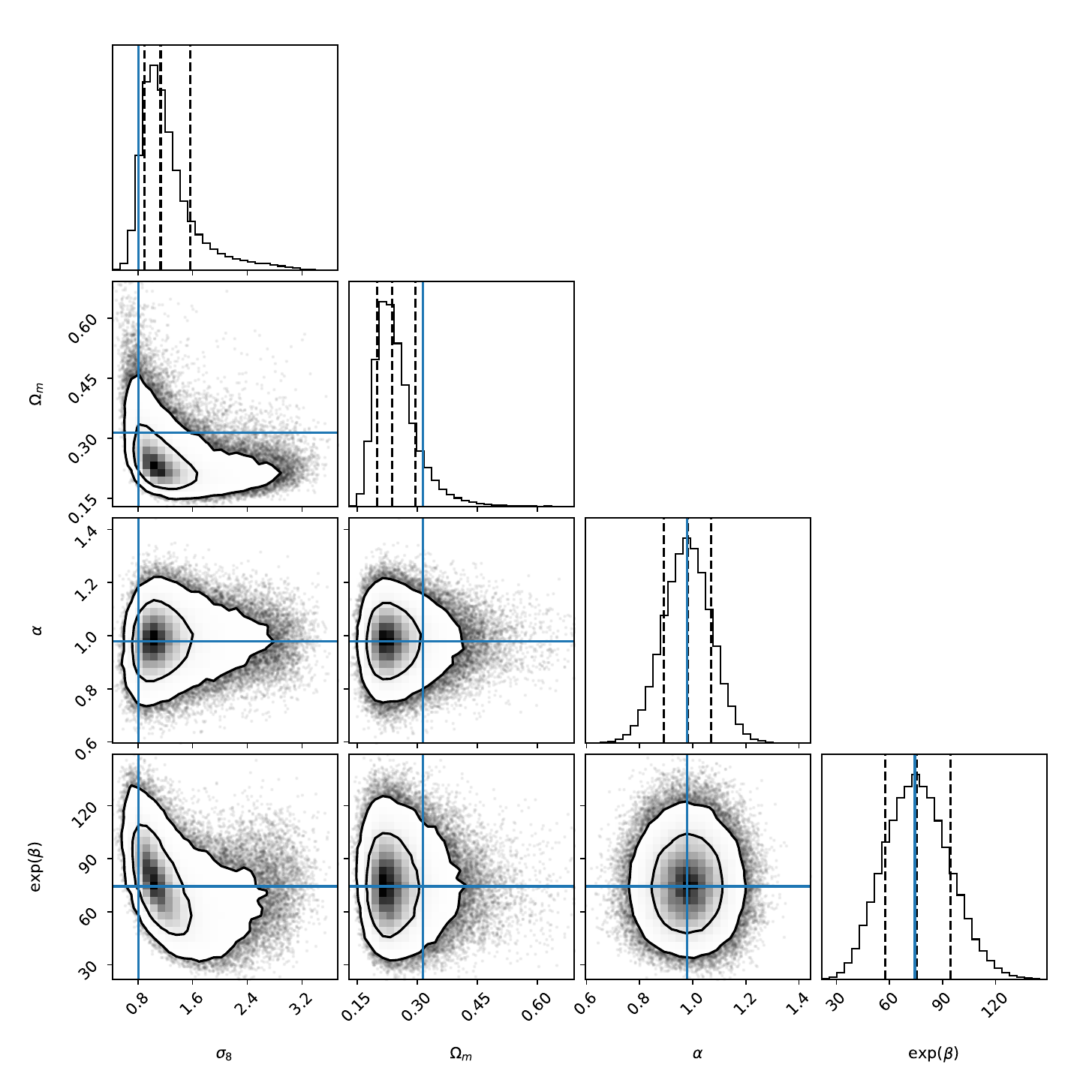}
        \caption{Posterior distribution for the two cosmological parameters $\Om$ and $\sigma_8$, and the two scaling relation parameters $\alpha$ and $e^{\beta}$. The 2D contours show the 68\% and 95\% confidence levels and the dashed lines in the 1D histograms show the 16\%, 50\%, and 84\% quantiles. The blue lines show the Planck 2018 values for $\Om$ and $\sigma_8$ and the K21 mean values for $\alpha$ and $e^{\beta}$.}
        \label{fig:mcmc}
    \end{figure}

\begin{figure}
        \centering
        \includegraphics[width=\hsize]{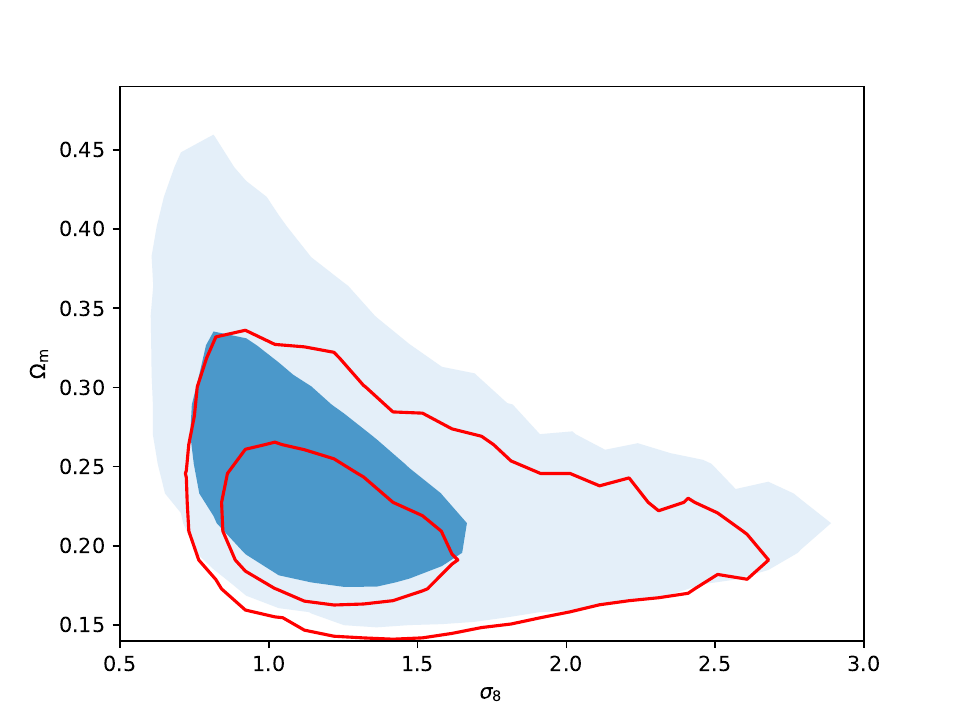}
        \caption{Posterior distribution for the two cosmological parameters, $\Om$ and $\sigma_8$, in the case of varying which wavenumbers are included in the sampling. The filled blue contours correspond to the case of $k < 0.2\hmpcinv$ and the red contours to the case of $k < 0.35\hmpcinv$.}
        \label{fig:mcmc_kmax}
    \end{figure}

    Figure~\ref{fig:mcmc} shows the posterior distributions for the two cosmological parameters $\Om$ and $\sigma_8$, along with the two scaling relation parameters, $\alpha$ and $e^{\beta}$, when we include all the three multipoles in the sampling. The plot was made using the \verb|corner|\footnote{\url{https://github.com/dfm/corner.py}} Python library \citep{foreman-mackey:2016}. From these, we obtained parameter constraints of $\Om = 0.24_{-0.04}^{+0.06}$ and $\sigma_8 = 1.13_{-0.24}^{+0.43}$, which correspond to the 16\%, 50\%, and 84\% quantiles of the marginalized posterior distributions. These can be compared, for example, to the Planck 2018 results, $\Om = 0.315 \pm 0.007$ and $\sigma_8 = 0.811 \pm 0.006$. Our marginalized constraints would suggest a larger than $1\sigma$ difference compared to the Planck 2018 results, but in Fig~\ref{fig:mcmc} the Planck 2018 value is indeed contained within the 68\% confidence region in the $(\sigma_8, \Om)$ plane. Regarding the richness-mass scaling relation, we obtained marginalized constraints of $\alpha = 0.98 \pm 0.09$ and $e^{\beta} = 75.5_{-17.8}^{+18.9}$. So, in essence, we reproduced the results of K21: $\alpha = 0.98 \pm 0.09$ and $e^{\beta}=74.4_{-18.2}^{+21.4}$. 
    
    Figure~\ref{fig:mcmc_kmax} shows the effect of including wavenumbers up to $0.35\hmpcinv$ in obtaining the posterior distribution for $\Om$ and $\sigma_8$. The results are statistically consistent with the $k < 0.2 \hmpcinv$ case. Going to smaller scales shrinks the $68\%$ and $95\%$ confidence regions by $36\%$ and $39\%$, respectively. Clearly, the additional data points at $0.2 \hmpcinv < k < 0.35 \hmpcinv$ increase the constraining power of the power spectrum measurement. However, due to potential problems with the covariance matrix estimate in this region (as demonstrated by Fig.~\ref{fig:pk_set}), we consider $k < 0.2 \hmpcinv$ our conservative baseline case.

\section{Conclusions}
\label{sec:conclusions}
    We presented power spectrum measurements for a subset of CODEX galaxy clusters and used these measurements to constrain cosmological parameters. This is a continuation of the work in L21, in which the cluster two-point correlation function was implemented. We have made some improvements to the L21 analysis methods, most notably regarding the covariance matrix estimation and the handling of the mass estimate uncertainties. 

    We found that the measured power spectrum multipoles are compatible with theoretical predictions that combine the linear matter power spectrum, the Kaiser approximation for redshift space distortions, and a large-scale bias estimated from the observed cluster richnesses. We constructed a likelihood function using the measured and predicted multipole signal and used this to generate constraints on the cosmological parameters $\Om$ and $\sigma_8$, which led to $\Om = 0.24_{-0.04}^{+0.06}$ and $\sigma_8 = 1.13_{-0.24}^{+0.43}$. These suggest slightly larger than $1\sigma$ deviations from Planck 2018 cosmology, for example. However, taking the degeneracy of these parameters into account shows that the 68\% confidence region of our posterior distribution does indeed contain the Planck 2018 cosmology.

    We see a few possible extensions as future work prospects. The full CODEX catalog consists of two disjoint patches. In this work, we included only one of them to be able to cover the sky region with mock catalogs. To expand our sample, we would need to use a set of mock catalogs that span larger cosmological volumes to cover the full CODEX footprint, or to cover the excluded patch with another set of mock catalog realizations drawn from the same ensemble as the ones used in this work. The latter option could cause an under-representation of large-scale correlations between the two patches, the importance of which should be verified. Another improvement would be to include the halo mass function in our MCMC sampling. Also, improving the modeling of the mock measurements would allow us to use the measured spectra up to larger wavenumbers in the cosmological analysis and most likely significantly tighten the obtained parameter constraints.

\begin{acknowledgements}
    The authors wish to acknowledge CSC – IT Center for Science, Finland, for computational resources.      
    It is a pleasure  to thank Isabella Baccarelli, Fabio Pitari, and Caterina Caravita for their support with the CINECA environment.The halo mocks were run on the Leonardo-DCGP supercomputer as part of the Leonardo Early Access Program (LEAP).
\end{acknowledgements}

\bibliographystyle{aa}
\bibliography{bibliography.bib}

\end{document}